\documentclass[sigconf]{acmart}
\AtBeginDocument{%
  }

\setcopyright{acmlicensed}
\copyrightyear{2018}
\acmYear{2018}
\acmDOI{XXXXXXX.XXXXXXX}
\acmConference[Conference XX]{Make sure to enter the correct
  conference title from your rights confirmation email}{June 03--05,
  2018}{Woodstock, NY}
\acmISBN{978-1-4503-XXXX-X/2018/06}

\usepackage{graphicx}
\usepackage{subcaption}
\usepackage{multirow}
\usepackage[ruled,vlined]{algorithm2e}
\usepackage{adjustbox}

\usepackage{tabularx}
\usepackage[table]{xcolor} 
\usepackage{array}
\usepackage{ragged2e} 
\usepackage{enumitem}

\newcolumntype{L}{>{\RaggedRight\arraybackslash}X} 

\begin{document}

\title{User Simulator-Guided Multi-Turn Preference Optimization for Reasoning LLM-based Conversational Recommendation}



\author{Xingyuan Xiang}
\authornote{These authors contributed equally to this work.}
\affiliation{%
  \institution{Huazhong University of Science and Technology}
  \city{Wuhan}
  \country{China}}
\email{taoistsword@gmail.com}

\author{Xiangchen Pan}
\authornotemark[1]
\affiliation{%
  \institution{Huazhong University of Science and Technology}
  \city{Wuhan}
  \country{China}}
\email{pxcstart666@gmail.com}

\author{WeiWei}
\authornote{Corresponding author.}
\affiliation{%
  \institution{Huazhong University of Science and Technology}
  \city{Wuhan}
  \country{China}
}
\email{weiw@hust.edu.cn}

\begin{abstract}
Conversational Recommender Systems (CRSs) leverage natural language interactions for personalized recommendation, yet information-scarce dialogue histories and single-turn recommendation paradigms may severely hinder accurate modeling of complex user preferences. To alleviate this issue, recent studies have introduced LLM-based user simulators, which generate natural language feedback and perform simulated multi-turn interactions to assist recommendation. Nevertheless, since simulators cannot access true user preference labels during inference, their feedback may deviate from actual user interests, causing errors to accumulate over multiple interactions and severely affecting the generalization of the recommender. Inspired by the multi-step reasoning capabilities of LLMs and the effectiveness of reinforcement learning in policy optimization, we propose SMTPO, a user simulator-guided multi-turn preference optimization conversational recommendation framework. To align simulator-generated feedback with true user preferences in the absence of explicit labels, we enhance feedback quality via multi-task supervised fine-tuning (SFT), enabling the simulator to better reflect users’ complex and diverse needs. To address the challenge of biased feedback destabilizing multi-turn optimization, we first allow the Reasoning LLM-based recommender to learn preference reasoning and recommendation patterns through SFT and then employ reinforcement learning with fine-grained reward design to progressively align with true user preferences, improving recommendation performance. Extensive experiments on public datasets demonstrate the effectiveness and transferability of our method.
\end{abstract}

\keywords{Conversational Recommendation, User Simulator, Multi-Turn Preference Optimization, Large Language Model, Reinforcement Learning}


\maketitle

\section{Introduction}
Conversational Recommender Systems (CRSs) aim to provide personalized recommendation to users via natural language interactions. The key challenge of CRS lies in the inherently brief nature of typical conversations, easily leading to insufficient historical context in modeling user preferences. As such, prior work often leverages external information (e.g., knowledge graphs (KGs)~\cite{chen2019towards,zhou2020improving} or reviews~\cite{zheng2024hycorec}) to enhance user preference understanding. Recently, Large Language Models (LLMs) have been integrated into CRS for their strong semantic understanding and reasoning capabilities, such as specially designed alignment strategies like zero-shot CRSs via prompting~\cite{he2023large}. Despite success, LLMs often struggle to effectively incorporate collaborative filtering (CF)~\cite{wu2024coral,zhu2024collaborative,zheng2024adapting}, a cornerstone of traditional recommendation systems. Accordingly, sequential research on LLM-driven CRS has increasingly focused on hybrid approaches to effectively overcome such issues, for instance, retrieval-reranking paradigm to identify similar candidates for LLM-based reranking~\cite{yang2024unleashing}, or retrieval-augmented generation to seamlessly integrate LLMs with CF for effectiveness validation~\cite{zhu2025collaborative}. Nonetheless, user preferences are often diverse and complex (even given rich knowledge), so relying solely on single-turn interaction may limit the accuracy of CRSs ~\cite{wang2025search}, while excessive engagement may severely degrade the overall conversational experience.

\begin{figure}[!tbp]
 \centering
  \includegraphics[width=0.5\textwidth]{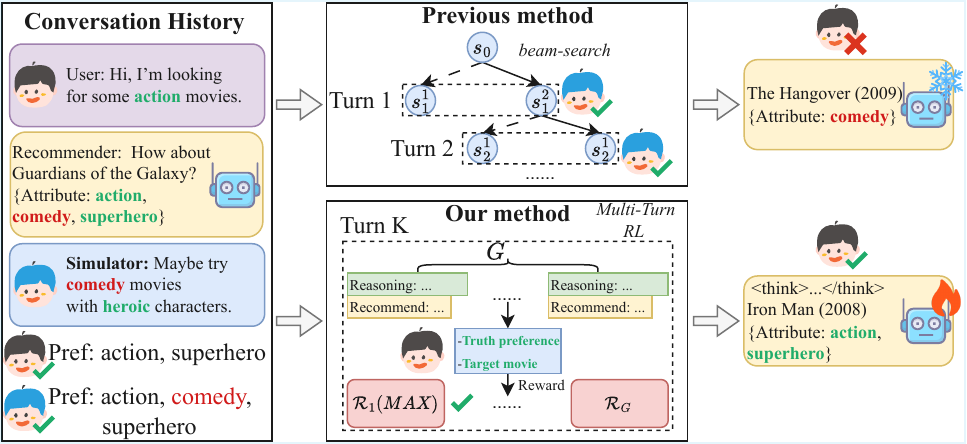}
  \caption{The previous method frozen recommender parameters, relied on beam search and simulator-filtered results, making it sensitive to feedback bias. Our method trains the SFT-initialized recommender via RL to achieve robust and accurate preference modeling and recommendation.}
  \label{fig:intro}
\end{figure}

To this end, recent works~\cite{feng2025expectation,yoon2024evaluating, wang2023rethinking} have explored leveraging LLM-based user simulators to enhance CRSs' understanding of complex user preferences and improve overall recommendation performance. These methods typically comprise two components, i.e., a \emph{user simulator} modeling true user preference and a \emph{recommender} generating context-aware recommendation based on dynamically generated user feedback, such as GRSU~\cite{wang2025search}, where a frozen LLM-based recommender produces candidates via beam search and is subsequently refined through a generative reward model-based simulator for optimal selection. However, it may be detrimental to the generalization of CRSs, as deviations between simulated user feedback and true user preferences can cause substantial error propagation. Without loss of generality, taking the case in Figure~\ref{fig:intro} as an example, the simulator incorrectly interprets the user’s preference as “comedy,” thereby biasing the recommender's filtering and leading the beam-search process to gradually deviate from the user’s actual preferences.

Actually, it is still non-trivial to design a proper LLM-based CRS framework, requiring solutions to two core challenges: (1) Effectively aligning simulator-generated feedback with true user preferences remains challenging, as ground-truth preferences are unavailable during inference, making the development of reliable preference-free user simulator a critical open problem. (2) Enhancing user preference modeling under biased or imperfect feedback is also difficult, as simulator-generated feedback may mislead the recommender, highlighting the need for the recommender to robustly achieve multi-turn preference optimization under dynamic and biased feedback.

To address these challenges, drawing on the success of practices of LLM-based agents~\cite{dong2024survey,huang2024understanding} as well as the strong multi-step reasoning capabilities of Reasoning LLMs~\cite{fang2025reason4rec,zhao2025reason}, we propose SMTPO (\textbf{S}imulator \textbf{M}ulti-\textbf{T}urn \textbf{P}reference \textbf{O}ptimization), a simulator-guided CRS framework based on Reasoning LLM. In SMTPO, the simulator generates high-quality feedback to guide recommendations, the retriever dynamically filters the candidate set, and the recommender uses both to iteratively optimize preferences and recommendations via reinforcement learning (RL). To tackle the challenge of generating high-quality feedback, we employ an LLM-based simulator to produce natural language responses and train it via multi-task supervised fine-tuning (SFT), enabling it to generate concrete and informative feedback that guides the recommender to make accurate recommendations. To ensure robust multi-turn preference optimization under biased feedback, we use a Reasoning LLM as the recommender backbone and adopt a two-stage training process: first, SFT is used to allow the recommender to initially grasp the task patterns; then, RL combined with fine-grained rewards is introduced, enabling the recommender to gradually align with true user preferences over multiple interactions, as illustrated in Figure~\ref{fig:intro}. Additionally, to prevent the LLM-based recommender from generating items outside the global item space~\cite{wang2025search}, we design a dual semantic-collaborative view retriever to dynamically constrain the candidate set and improve recommendation accuracy.

In summary, the main contributions of this paper are as follows:
\begin{itemize}[leftmargin=*] 
    \item We propose a novel multi-turn interactive conversational recommendation framework, SMTPO. To the best of our knowledge, this is the first work that jointly trains the recommender via SFT and RL, enabling continuous optimization of user preferences across multiple interaction turns.
    \item We construct a simulator-guided CRSs consisting of a user simulator, a retriever, and a recommender. Multi-task fine-tuning improves the quality of simulator-generated feedback, and the retriever effectively integrates feedback with the candidate set, thereby stably enhancing recommendation performance.
    \item Extensive experiments further demonstrate that our method significantly outperforms existing baselines on multiple datasets, verifying its effectiveness and superiority.
\end{itemize}

\section{Related Work}
\textbf{LLM-based Conversational Recommender Systems (CRSs).} Existing CRS research can be broadly categorized into attribute-based CRSs~\cite{deng2021unified,xu2021adapting,lei2020estimation,lei2020interactive} and generation-based CRSs~\cite{wei2025mscrs,dao2024broadening,li2023trea,wang2022towards}. Attribute-based CRSs rely on fixed templates, leading to mechanical interaction patterns, while generation-based CRSs combine pre-trained language models (PLMs) with knowledge graphs but still struggle to model diverse user preferences. Given the powerful semantic understanding capabilities of LLMs~\cite{wang2024can,wei2022chain}, recent studies have introduced them into CRS to address these limitations. For example, ~\citet{he2023large} systematically analyzed the performance of LLMs in zero-shot CRS; ~\citet{xi2024memocrs} proposed MemoCRS, which leverages a memory-augmented LLM to manage users’ historical preferences, thereby enhancing the personalization of recommendations; ~\citet{yang2024unleashing} unified the LLM into a CRS with both retrieval and generation capabilities through instruction tuning. LLMs not only possess strong language understanding and generation capabilities but can also exhibit multi-step reasoning abilities through reinforcement learning (RL)~\cite{yu2025dapo,yue2025vapo} or multi-path sampling~\cite{brown2024large,wang2022self}. RL-based post-training methods, such as DeepSeek-R1~\cite{guo2025deepseek} and Kimi k1.5~\cite{team2025kimi}, enable user preference analysis and candidate item evaluation. Although the reasoning capabilities of LLMs have been validated in structured tasks, their potential in CRS remains largely unexplored.

\textbf{User Simulator for CRSs.} Although LLMs bring new opportunities to CRSs, they often struggle to model user preferences with limited context, and frequent reliance on real users increases interaction costs. To address this, prior studies introduced user simulators that generate natural language feedback to assist recommender optimization~\cite{feng2025expectation,yoon2024evaluating, wang2023rethinking,lin2024interpretable}. Early works mainly used simulators for evaluation with ground-truth preference labels~\cite{yoon2024evaluating, wang2023rethinking}, making them difficult to apply directly in real interaction scenarios. ~\citet{wang2025search} were the first to propose a simulator that does not require preference labels, enabling automatic interaction with CRSs. However, due to the lack of a proper supervision mechanism, these methods become unreliable when the simulator’s feedback is biased. The generalization ability of the recommender still needs improvement.

Our method belongs to the LLM-based CRS and introduces a user simulator. Unlike previous methods, SMTPO employs a label-free simulator to generate high-quality natural language feedback and adopts a Reasoning LLM as the recommender backbone, leveraging its strong multi-step reasoning ability and RL algorithm, SMTPO achieves multi-turn preference optimization to filter noisy feedback, and significantly enhances recommendation performance.


\section{Methodology}
\begin{figure*}[ht]
  \includegraphics[width=\textwidth]{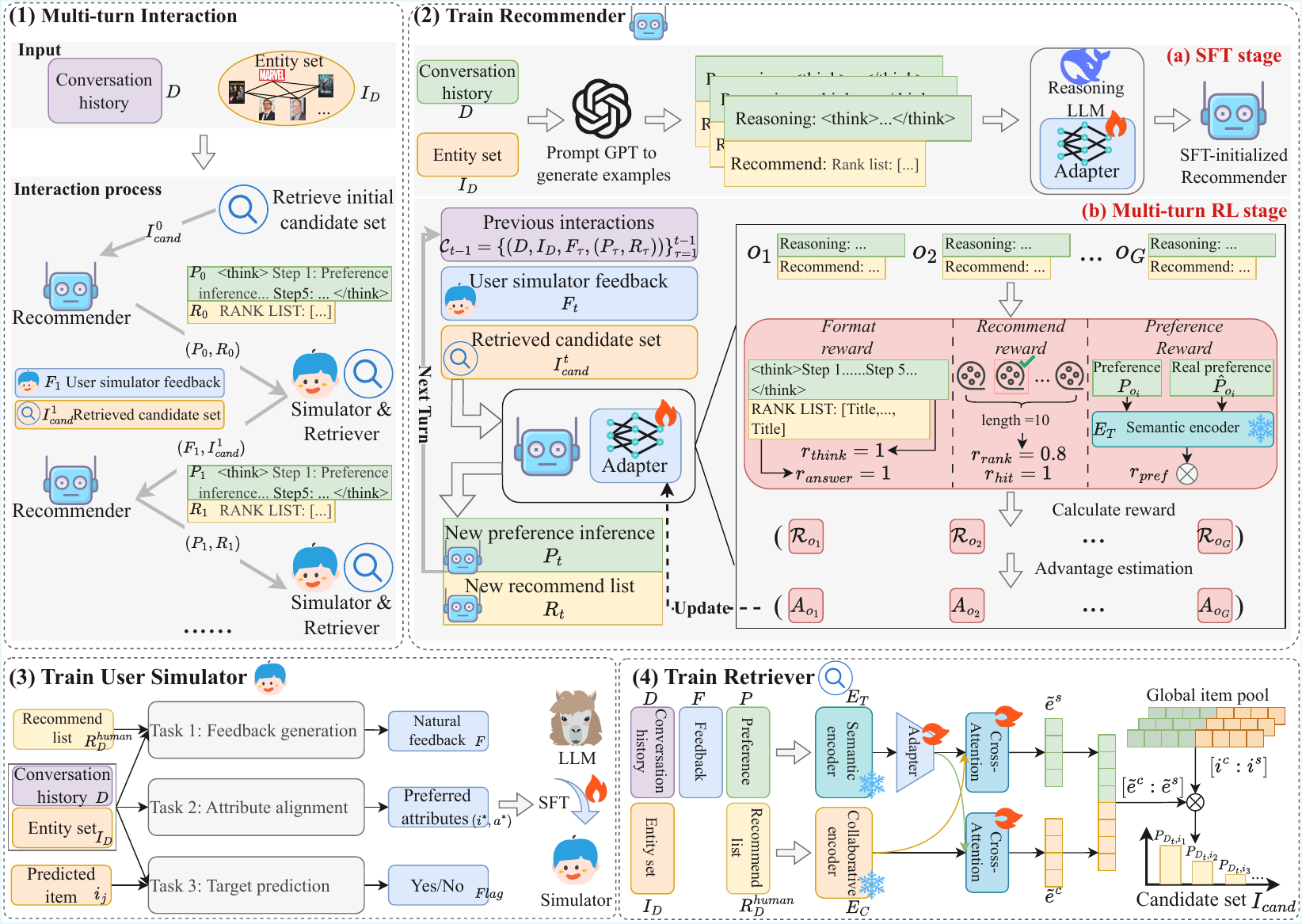}
  \caption{Overview of SMTPO training and multi-turn interaction: (1) The multi-turn interaction process among the recommender, simulator, and retriever. (2) The recommender is trained first with SFT and then with multi-turn RL. (3) The simulator is trained via multi-task SFT. (4) The retriever is obtained using collaborative–semantic dual-view modeling.}
  \label{fig:method}
\end{figure*} 
\subsection{Overview of SMTPO}
In this paper, we propose a multi-turn preference optimization framework for conversational recommendation, named \textbf{SMTPO}, which is composed of three core modules: \textbf{user simulator}, \textbf{retriever} and the Reasoning LLM-based \textbf{recommender}. The overall framework is illustrated in Figure~\ref{fig:method}.  

\textbf{Task formulation:} Conversational recommender systems (CRSs) aim to provide recommendations through multi-turn interactions. Considering that a single-turn interaction cannot fully capture complex and diverse user preferences, in this work, we introduce a label-free user simulator to provide feedback and interact automatically with the recommender for more personal recommendation. The user simulator generates feedback $F_t$ based on the conversation history $D$, the set of entities and attributes $I_D=\{(i_1, a_1), (i_2, a_2), \dots, (i_n, a_n)\}$ mentioned in the conversation, and the previous turn recommendation results $R_{t-1}$. After receiving the feedback, the recommender performs preference inference $P_t$ and outputs new recommendation results $R_t$. Formally, the multi-turn interaction between the user simulator and the recommender up to the turn $t$ is represented as the sequence:
\begin{equation}
\mathcal{C}_t = \{ (D, I_{D}, F_\tau, (P_\tau, R_\tau)) \}_{\tau=1}^t.
\end{equation}
Based on the above definitions, the CRS task under multi-turn interaction can be formalized as:
\begin{equation}
\begin{cases}
F_t = \text{Simulator}(D, {I}_D, R_{t-1}), \\
(P_t, R_t) = \text{Recommender}(D, {I}_D, F_t),
\end{cases} \quad t = 1, \dots, T,
\end{equation}
where $R_0$ denotes the initial recommendation list. The interaction proceeds iteratively until reaching the maximum number of turns $T$ or the target item $i^\ast$ is successfully recommended.

\textbf{Retrieval-reranking paradigm:} To prevent the LLM-based recommender from generating results over the global item space and motivated by the proven effectiveness of the retrieval\--reranking paradigm in prior work~\cite{yang2024unleashing,zhu2025collaborative}, we introduce a retriever between the simulator and the recommender to provide a candidate item set. Specifically, we first leverage the simulator-generated user feedback $F_t$ and the recommender’s previous preference inference $P_{t-1}$, together with the conversation history $D$ and entity information $I_D$, to jointly model the representation of the current interaction from semantic and collaborative perspectives. Formally, at turn $t$, the semantic side information consists of the conversation history $D$, user feedback $F_t$, and user preferences $P_{t-1}$; the collaborative side information is represented by the collaborative embeddings of both the entity set $I_D$ and the previous recommendation list $R_{t-1}$ within the knowledge graph $\mathcal{G}$. The retriever computes the interaction representation $d_t$ and calculates its similarity with all item embeddings to obtain the candidate item set $I_\text{cand}^t$.

\textbf{Module training:} To ensure more stable multi-turn interaction training, the simulator and retriever modules are trained independently and prior to the recommender module. The simulator is trained to generate high-quality user feedback, achieved via multi-task supervised fine-tuning (SFT), as detailed in \autoref{subsec:simulator}. The retriever is trained to recall candidate item sets containing the target items, realized through collaborative-semantic dual-view modeling, as detailed in \autoref{subsec:retriever}. The recommender is trained to continuously and stably optimize preference understanding and improve recommendation performance during multi-turn interactions. We use the reinforcement learning (RL) algorithm to achieve this, as detailed in \autoref{subsec:recommender}.

\subsection{User Simulator: Generate High-quality Feedback}
\label{subsec:simulator}
In our approach, the user simulator is designed to provide user feedback to the recommender based on the current recommendation list and the conversation history, enabling the recommender to update its understanding of user preferences and generate more accurate recommendations. However, providing high-quality user feedback in the absence of true preference labels remains a significant challenge. To address this, we have designed a multi-task SFT method to train the LLM-based simulator, ensuring that the feedback generated by the simulator resembles real user behavior. This section describes the task design and training procedure of the simulator.

\subsubsection{Feedback Generation Task}
Considering that the goal of the user simulator is to provide specific and informative user feedback, we design a feedback generation task that enables the simulator to generate feedback based on conversation history $D$ and the items and attributes $I_D$ mentioned in the conversation. The instruction input includes $D$, $I_D$. To better reflect realistic multi-turn interaction scenarios, we also include a human-constructed recommendation list $R_D^{\text{human}}$ in the input. To cover as many interaction cases as possible, we design four construction strategies: (1) hard negative sample list containing the target item, (2) hard negative sample list excluding the target item, (3) simple negative sample list containing the target item, and (4) simple negative sample list excluding the target item, denoted as $H^+$, $H^-$, $S^+$, and $S^-$. The difficulty of the negative samples is measured by the overlap of their attributes with those of the target item. We employ GPT-3.5-Turbo~\cite{shafik2024introduction} to construct high-quality feedback as supervision signals, where the information of the target item is additionally included in the instruction to ensure the correctness of the ground truth. The formal definition of the feedback generation task is presented as follows:
\begin{equation}
F = LLM_{simulator}(D, I_D, R_D^{\text{human}})
\end{equation}
where $R_D^\text{human} \in \{H^+, H^-, S^+, S^-\}$.

\subsubsection{Attribute Alignment Task}
To ensure that the simulator generates high-quality feedback, we require the simulator to resemble real users not only in style and semantics but also in alignment with the target item at the attribute level for finer-grained alignment with user preferences. Thus, we design an attribute alignment task, where the simulator predicts the target item $i^\ast$ and its attributes $a^\ast$ based on the conversation history $D$ and mentioned entity set $I_D$. The formal definition is as follows:
\begin{equation}
\begin{aligned}
(i^\ast, a^\ast) = LLM_{simulator}(D, I_D)
\end{aligned}
\end{equation}

\subsubsection{Target Prediction Task}
During multi-turn interactions, the recommendation list generated by the recommender may contain noise. Specifically, some candidate items may have semantics inconsistent with the conversation history or the user’s true preferences. The simulator requires strong judgment to filter out these irrelevant items, so we designed a target prediction task. Let the simulator determine whether the current item $i_j$ (where $i_j \in H^+ \cup H^- \cup S^+ \cup S^-$) is the target item of the user, based on the dialogue history $D$ and the entity set $I_D$. The simulator should return a variable $Flag$ ("Yes" or "No") as its response. The formal definition is as follows:
\begin{equation}
\begin{aligned}
Flag = LLM_{simulator}(i_j|D, I_D)
\end{aligned}
\end{equation}

\textbf{Model training:} We adopt LoRA~\cite{hu2022lora} to perform multi-task SFT on the LLM to obtain the user simulator. During training, for each task instance $(\mathcal{I}, o) \in \text{Inst}$, the model updates the parameters $\Theta^u$ by maximizing the conditional probability of the target output sequence. The loss function is defined as follows:
\begin{equation}
\begin{aligned}
\mathcal{L}_{simulator}=-\sum_{(\mathcal{I}, o) \in Inst}\sum_{i=1}^{|o|}logPr(o_i|o_{<i}, \mathcal{I};\Theta^u)
\end{aligned}
\end{equation}
Here, $\mathcal{I}$ denotes the input instruction and $o$ denotes the output.

\subsection{Retriever: Recall Candidate Set}
\label{subsec:retriever}
To avoid the LLM-based recommender generating items outside the global item space, we adopt a retrieval–reranking paradigm and design a dual-view retriever. It encodes each dialogue from semantic and collaborative perspectives to retrieve a candidate set from the global item pool.

\subsubsection{Feature Embedding}
\textbf{Semantic Embedding}: On the semantic side, we use a pre-trained language model (PLM) as the semantic encoder $E_T$, which encodes the conversation history $D$, user feedback $F$, and the recommender’s textual preference inference $P$ to obtain semantic representation $e^s$:
\begin{equation}
\begin{aligned}
e^s = E_T(D, F, P)
\end{aligned}
\end{equation}

\textbf{Collaborative Embedding}: 
Besides semantic information, structural relations among entities provide rich collaborative filtering knowledge. We pretrain an encoder $E_C$ to model this knowledge by constructing a knowledge graph $\mathcal{G}$ from WikiMKG~\cite{qiu2024knowledge} and applying a GCN model for entity representation. To capture the potential collaborative signals between entities and their related attributes, we treat each movie–attribute pair as a positive sample and unrelated attributes as negative samples and optimize $E_C$ using the BPR loss. After pretraining, we freeze $E_C$, encode $I_D$ and the human-constructed recommendation list $R_D^{human}$, using mean pooling to obtain collaborative representation $e^c$:
\begin{equation}
\begin{aligned}
e^c = AvgPool(E_C(I_D,R_D^{human}))
\end{aligned}
\end{equation}

\subsubsection{Dual-view Modeling}
After obtaining the initial features, in order to better integrate the conversation representations of the two views, we adopt a cross-attention mechanism for feature fusion. Due to the semantic dimension $Dim_s$ and collaborative dimension $Dim_c$ not being the same, spatial alignment is required first. We use an adapter to map semantic information $e^s$ to the collaborative space. This process can be formally defined as follows:
\begin{equation}
\begin{aligned}
\hat{e}^s = W_2(W_1 e^s + b1) + b2
\end{aligned}
\end{equation}
Where $W_1 \in R^{Dim_s \times \frac{Dim_s}{2}}$, $W_2 \in R^{\frac{Dim_s}{2} \times Dim_c}$, $b_1 \in R^{\frac{Dim_s}{2} \times 1}$, $b_2 \in R^{Dim_c \times 1}$ are weight matrices and bias of the adapter.

After completing spatial alignment, feature fusion is performed based on cross-attention. Taking the collaborative side as an example, we set it $\hat{e}^s$ as the query vector and $e^c$ as the key and value vector. Let $Q_s=\hat{e}^s W^Q$, $K_c=e^c W^K$,$V_c=e^c W^V$, where $W^Q$,$W^K$,$W^V$ $\in R^{Dim_c \times Dim_c}$, the collaborative feature that integrates semantic side information can be represented as:
\begin{equation}
\begin{aligned}
\tilde{e}^c = Softmax(\frac{Q_sK_c^T}{\sqrt{D}})V_c
\end{aligned}
\end{equation}

The same approach can be used to obtain semantic features that integrate collaborative side information $\tilde{e}^s$. By concatenating these two fusion vectors, the feature representation of conversation $D$ can be obtained $e = [\tilde{e}^c:\tilde{e}^s]$. Similarly, we concatenate the text features $i^s$ and collaborative features $i^c$ of the item as the fused features on the item side. We then retrieve the recommended candidate set by calculating feature similarity. For the given conversation $D$ and item $i$, the recall probability is calculated as follows:
\begin{equation}
\begin{aligned}
P_{D,i} = [\tilde{e}^c:\tilde{e}^s]^T[i^c:i^s]
\end{aligned}
\end{equation}
where $i^s$ is encoded by the text encoder $E_T$ based on the text description of the item, and $i^c$ is encoded by the graph embedding model $E_C$. Finally, we rank and return the top-k items by recall probability as a candidate item set for conversation $D_t$ in the t-th turn of interaction.

\textbf{Model training}: As mentioned earlier, we obtain collaborative features by a GCN model, which is pre-trained based on $\mathcal{G}$. Regarding the data construction on the text side of the training set, to better simulate real interaction scenarios as closely as possible, user feedback was generated by the user simulator trained in~\autoref{subsec:simulator}, and user preferences were simulated using GPT-3.5-Turbo during the data construction stage. In addition, to ensure stable and efficient training, the parameters of $E_T$ and $E_C$ are frozen during the retriever training phase. Based on the final recall probability, we use InfoNCE loss to train the retriever module:
\begin{equation}
\begin{aligned}
\mathcal{L}_{retrieval} = -log\frac{exp(P_{D, i^{\ast}})}{exp(P_{D, i^{\ast}}) + \sum_{k=1}^N exp(P_{D, i_{neg}^k})}
\end{aligned}
\end{equation}
where $i^{\ast}$ represents the target item of conversation $D$, $i_{neg}$ is the negative sample, and $N$ is the number of negative samples. Negative samples are obtained by randomly sampling after excluding the target item and the entities mentioned in the dialogue.
                    
\subsection{Recommender: Reason User Preferences and Fine-grained Ranking}
\label{subsec:recommender}
The role of the recommender is to analyze user preferences based on the conversation history and the user feedback, and to rerank the candidate set. Due to the excellent language comprehension and reasoning capabilities of the Reasoning LLM, we use it as the foundation of the recommender and optimize it through a two-stage training paradigm.

\subsubsection{Two Stage Model Training}
We propose a two-stage model training approach. In the first stage, we perform SFT to help the model develop an initial understanding of the reasoning process and the recommendation task. In the second stage, we apply multi-turn RL to enhance the model’s deep reasoning and adaptive adjustment capabilities during multi-turn interactions, ultimately achieving more robust recommendation performance.

\textbf{SFT stage:} We first define the interaction scenario and task setup of the recommender. The instruction input $Q$ consists of the task description $Inst$, dialogue history $D$, candidate item set $I_{cand}$, and user feedback $F$. The candidate set $I_{cand}$ is obtained from the retriever module, while the task description $Inst$ requires the recommender to analyze user preferences based on $\{D, I_{cand}, F\}$ and re-rank the candidate items, with the output being the Top-10 item list $R$. To enhance the recommender’s capability for deep reasoning, we further require the model to follow a step-by-step reasoning process before providing the final answer, including preference inference → attribute matching → scoring → ranking → recommendation explanation. The gold reasoning trajectory $\hat{T}$ and recommendation results $\hat{R}$ are generated by GPT-3.5-Turbo, with additional inputs $R_D^\text{human} \in {H^+, H^-, S^+, S^-}$ and the user’s target item $i^{\ast}$ and attribute $a^{\ast}$. Thus, each training sample is constructed as a triplet $(Q,T,R)$, where $T$ denotes the reasoning trajectory and $R$ denotes the recommendation output. The fine-tuning objective is formally defined as follows:
\begin{equation}
\begin{aligned}
\theta^* = \arg\min_\theta \mathbb{E}_{(Q,T,R)\sim \mathcal{D}_{sft}} \mathcal{L}_{sft}(LLM_\theta(Q),T,R)
\end{aligned}
\end{equation}
where $\mathcal{D}_{sft}$ represents the training dataset in the SFT stage and $\mathcal{L}_{sft}$ is used to encourage the recommender to generate reasoning chains and answers consistent with standard supervised data.

\textbf{Multi-turn RL stage:} To better align with real multi-turn interaction scenarios and enhance the recommender’s reasoning over user preferences, we introduce a multi-turn RL stage following the SFT stage. Unlike single-turn supervised training, multi-turn interactions continuously accumulate historical information and user feedback, where biased feedback from the simulator can lead to significant error accumulation. To ensure stable preference optimization under such bias, we employ the GRPO algorithm to model the recommender’s policy evolution across multi-turn interactions, with the optimization objective defined as:
\begin{equation}
\begin{aligned}
\mathcal{J}_{GRPO}(\theta) = 
& \; \mathbb{E}_{q\sim P(Q),\{o_i\}_{i=1}^G \sim \pi _{\theta_{\text{old}}}(O|q)} \Bigg[ \frac{1}{G}\sum_{i=1}^G 
    \min\Bigg( 
        \frac{\pi_\theta(o_i|q)}{\pi_{\theta_{\text{old}}}(o_i|q)} A_i, \\
& \quad \operatorname{clip}\!\Bigg( 
        \frac{\pi_\theta(o_i|q)}{\pi_{\theta_{\text{old}}}(o_i|q)}, 
        1-\epsilon, 1+\epsilon 
    \Bigg) A_i 
\Bigg) 
- \beta D_{KL}(\pi_\theta \,\|\, \pi_{\text{ref}})
\Bigg] 
\end{aligned}
\end{equation}
where $A_i$ represents the relative advantage within the group. By normalizing the action rewards within the same group, stable training can be achieved and efficiency can be improved.

\subsubsection{Reward Design}
In order to provide stable and effective guidance for multi-turn optimization of the model, we have designed three types of reward functions, namely format reward, recommend reward, and preference reward. Next, we describe the reward computation process at each turn $t$.

\textbf{Format reward:} The format reward is mainly used to encourage models to reason step-by-step according to the specified reasoning procedure and then provide answers after reasoning. Here, we design corresponding reward functions for both the reasoning process and the answer to impose format constraints.

Process format reward: For the process format reward, we mainly consider whether the model reasoning steps comply with the five-step specification (preference inference, attribute matching, scoring, sorting, explanation). By performing regular matching on the <think> labels output by the model, the score is calculated based on the number of matches $N_{match}$. The reward function is as follows: 
\begin{equation}
\begin{aligned}
r_{think} &= \left\{
\begin{array}{ll}
1 - \frac{|N_{match}-5|}{5}, & \text{if } 1 \leq N_{match} \leq 7 \\
0, & otherwise
\end{array}
\right. 
\end{aligned}
\end{equation}

Answer format reward: As the recommendation task is a reranking task and returns a top-k item list, we determine whether it matches the standard answer format $Ans$: (RANK LIST: [Title,..., Title]). The answer reward function is shown as follows:
\begin{equation}
\begin{aligned}
r_{answer} &= \left\{
\begin{array}{ll}
1, & \text{if match ($Ans$)} \\
0, & \text{otherwise}
\end{array}
\right. 
\end{aligned}
\end{equation}

\textbf{Recommend reward:} This reward is used to enhance the recommendation ranking ability of the model. Based on the target item $i^{\ast}$ in the conversation, the overall quality of the refined ranking result $R_t$ provided by the model is evaluated. Here, we design two metrics, Hit and Rank.

Hit reward: Hit reward is used to determine whether the target item is on the refined ranking list, formally defined as follows:
\begin{equation}
\begin{aligned}
r_{hit} = \left\{
\begin{array}{ll}
1, & \text{if } i^{\ast} \in R_t \\
0, & \text{otherwise}
\end{array}
\right. 
\end{aligned}
\end{equation}

Rank reward: Rank reward is a linear decay reward. Based on the position of the target item $i^{\ast}$ at $pos_{i^{\ast}} \in [1, len(R_t)]$ in the refined ranking list $R_t$, the formal definition is as follows:
\begin{equation}
\begin{aligned}
r_{rank} = \left\{
\begin{array}{ll}
1-\frac{pos_{i^{\ast}}-1}{\mathrm{len}(R_t)}, & \text{if } i^{\ast} \in R_t \\
0, & \text{if } i^{\ast} \notin R_t
\end{array}
\right. 
\end{aligned}
\end{equation}

\textbf{Preference reward:} The preference reward is used to enhance the model's preference understanding ability. We extract the preference inference $P_t$ from the reasoning chain and compare it with the standard preference of the conversation $\hat{P}$ and calculate semantic similarity to determine whether the current model's preference understanding is accurate.
\begin{equation}
\begin{aligned}
r_{Prefer} = f_{sim}(E_T(P_t), E_T(\hat{P}))
\end{aligned}
\end{equation}
Where $f_{sim}$ is the cosine similarity function, and $E_T$ is the text encoder.

\subsection{Multi-Turn Interaction}
This section introduces the multi-turn interaction process of SMTPO. As shown in  Figure~\ref{fig:method}, in the first turn, the retriever retrieves an initial candidate set $I_{\text{cand}}^{0}$ based on the conversation history $D$ and the entity set ${I}_D$. The recommender then performs preference reasoning $P_0$ and outputs the initial recommendation list $R_0$. In the subsequent turn $t$, the simulator generates feedback $F_t$ based on the previous recommendation list. The retriever combines the current feedback with the previous preference to retrieve a new candidate set $I_{\text{cand}}^{t}$. The recommender then updates preference reasoning $P_t$ and generates a new recommendation list $R_t$ based on the previous interactions $\mathcal{C}_t = \{ (D, I_{D}, F_\tau, (P_\tau, R_\tau)) \}_{\tau=1}^t$. The process terminates when the maximum number of turns $T$ is reached or the target item $i^\ast$ is successfully recommended.

\section{Experiments}
\begin{table*}[ht]
  \centering
  \Huge
  \caption{Comparison of the main results between our method and baseline methods. The best result is given in bold. Significant improvements are marked with * (t-test, p < 0.05).}
  \label{tab:recommendation}
  \renewcommand{\arraystretch}{1.5}
  \begin{adjustbox}{max width=\textwidth}
  \begin{tabular}{c c | c c c c c c c c c c c c c c}
    \toprule
    Dataset & Metric 
      & Popularity & BERT 
      & L3.1-8B-I & GPT-3.5 & GPT-4o 
      & ReDial & KBRD & KGSF & TREA 
      & VRICR & UniCRS & DCRS & MSCRS & \bf{SMTPO} \\
    \midrule
    \multirow{4}{*}{ReDial} 
      & Recall@1  & 0.011  & 0.027 & 0.037 & 0.043 & 0.047 & 0.010 & 0.033 & 0.035 & 0.045 & 0.054 & 0.065 & 0.076 & \underline{0.081} & \bf{0.092*} \\
      & Recall@10 & 0.053  & 0.142 & 0.136 & 0.165 & 0.189 & 0.065 & 0.150 & 0.175 & 0.204 & 0.244 & 0.241 & 0.253 & \underline{0.264} & \bf{0.270*} \\
      & NDCG@10   & 0.029 & 0.075 & 0.067 & 0.091 & 0.106 & 0.034 & 0.083 & 0.094 & 0.114 & 0.138 & 0.143 & 0.154 & \underline{0.161} & \bf{0.165*} \\
      & MRR@10    & 0.021 & 0.055 & 0.046 & 0.069 & 0.077 & 0.024 & 0.062 & 0.070 & 0.087 & 0.106 & 0.113 & 0.123 & \underline{0.128} & \bf{0.130*} \\
    \midrule
    \multirow{4}{*}{INSPIRED} 
      & Recall@1  & 0.031 & 0.049 & 0.043 & 0.053 & 0.079 & 0.009 & 0.042 & 0.051 & 0.047 & 0.043 & 0.085 & 0.093 & \underline{0.096} & \bf{0.099*} \\
      & Recall@10 & 0.155 & 0.189 & 0.162 & 0.177 & 0.196 & 0.048 & 0.135 & 0.132 & 0.146 & 0.141 & 0.230 & 0.226 & \underline{0.257} & \bf{0.273*} \\
      & NDCG@10   & 0.085 & 0.112 & 0.097 & 0.109 & 0.121 & 0.023 & 0.088 & 0.092 & 0.095 & 0.091 & 0.149 & 0.153 & \underline{0.168} & \bf{0.174*} \\
      & MRR@10    & 0.064 & 0.088 & 0.077 & 0.088 & 0.098 & 0.015 & 0.073 & 0.079 & 0.076 & 0.075 & 0.125 & 0.130 & \underline{0.140} & \bf{0.142*} \\
    \bottomrule
  \end{tabular}
  \end{adjustbox}
\end{table*}

In this section, we attempt to answer the following research questions (RQs):
RQ1: How does our method perform on the overall conversational recommendation task? (Sec ~\ref{ex_RQ1})
RQ2: Does the training process of the user simulator and the recommender help improve recommendation performance? (Sec ~\ref{ex_RQ2})
RQ3: Does the multi-turn preference optimization process lead to improvements in recommendation performance? (Sec ~\ref{ex_RQ3})
RQ4: Where do the gains in recommendation performance come from, and does the system truly elicit user preferences? (Sec ~\ref{ex_RQ4})


\subsection{Experiment Setups}
\subsubsection{Dataset}
To verify the effectiveness of our method, we conduct experiments on two widely used  CRS datasets, namely ReDial~\cite{li2018towards} and INSPIRED~\cite{hayati2020inspired}. ReDial contains 10,006 dialogues, 956 users, and 6,924 items, while INSPIRED is slightly smaller, containing 10,01 dialogues, 1,482 users, and 1,123 items. We follow the same dataset partitioning strategy as in the previous works.

\subsubsection{Baselines}
We compare with the following baselines:

(1) Conventional recommendation models: We include Item Popularity, which ranks items by their historical frequency, and BERT~\cite{devlin2019bert}, a PLM fine-tuned to predict candidate items.

(2) General LLMs as Zero-shot CRSs: We evaluate LLMs in a zero-shot setting, including Llama-3.1-8B-Instruct~\cite{dubey2024llama} (L3.1-8B-I), an open-source 8B-parameter model, and GPT-3.5-Turbo and GPT-4o~\cite{achiam2023gpt}, which are closed-source models developed by OpenAI.

(3) State-of-the-Art CRS Methods: This group includes ReDial~\cite{li2018towards},  KBRD~\cite{chen2019towards}, a knowledge-enhanced CRS using subgraphs from DBpedia; KGSF~\cite{zhou2020improving}, leveraging item and word oriented KGs; TREA~\cite{li2023trea}, employing tree-structured reasoning; VRICR~\cite{zhang2023variational}, adopting variational bayesian pretraining; UniCRS~\cite{wang2022towards}, unifying recommendation and generation via prompt tuning; DCRS~\cite{dao2024broadening}, retrieving contextually similar dialogues to enhance recommendation; and MSCRS~\cite{wei2025mscrs}, integrating collaborative and multimodal information through semantic graphs and prompt learning.

\subsubsection{Metric}
Given the remarkable performance of LLMs in conversation tasks~\cite{chang2024survey}, we follow existing work~\cite{xie2024neighborhood,wang2025search} to focus our evaluation primarily on the recommendation task. We use Recall@k, NDCG@k, and MRR@k for evaluation, with $k \in \{1, 10\}$. We clarify that during inference, each dialogue instance in the test set is independently run for the full multi-turn interaction (up to five turns). For each turn $t$, recommendation metrics are computed independently on the same fixed and complete test set, rather than on the subset of dialogues that have not yet reached the target.

\subsubsection{Implementation Details} 
For user simulator, we adopt Llama-3.1-8B-Instruct as the backbone. The model is SFT-trained for 5 epochs with a learning rate of $1\times10^{-5}$, LoRA rank 8, and the AdamW optimizer.
For retriever, training runs up to 50 epochs with early stopping, a learning rate of $5\times10^{-5}$, and the Adam optimizer. The top-20 candidates are retrieved using bge-base-en-v1.5~\cite{xiao2024c} for text encoding and LightGCN~\cite{he2020lightgcn} for collaborative encoding.
For recommender, we adopt DeepSeek-R1-Distill-Llama-8B~\cite{guo2025deepseek}. It is SFT-trained for 8 epochs (learning rate $5\times10^{-5}$, LoRA rank 8, AdamW), followed by multi-turn GRPO optimization for 3 epochs. The group size of GRPO is 4, with gradient accumulation 8, max length 2048, learning rate $1\times10^{-5}$, and 5 interaction turns. FlashAttention2~\cite{dao2023flashattention} accelerates training. The recommender receives 20 candidates and outputs a final top-10 list. 
\subsection{Overall Performance Analysis}
\label{ex_RQ1}
To answer RQ1, we compare SMTPO with baselines in Table~\ref{tab:recommendation} and summarize the following observations:

(1) Traditional methods perform poorly. BERT outperforms Popularity but remains far below most methods, showing that semantic understanding alone cannot capture complex user preferences.

(2) General LLMs are competitive but not optimal. Zero-shot LLMs underperform specialized CRS models, highlighting the need for CRS-specific optimization.

(3) Representative CRS baselines perform well, each with distinct strengths. Notably, UniCRS and DCRS excel by combining entity representations with dialogue context and prompt-based optimization, while DCRS outperforming UniCRS via contextual retrieval. MSCRS integrates multi-modal features and semantic graphs, showing that enriching contextual representation with multi-source information is effective and motivates further use of simulated feedback and multi-turn preference optimization.

(4) Our proposed SMTPO outperforms all baselines, thanks to its design for feedback effectiveness and stable preference modeling. The simulator, fine-tuned with multi-task instructions, generates high-quality feedback without true labels, mitigating failure and misleading issues during the inference stage. The recommender undergoes two-stage training, first with SFT and then with RL, where during the multi-turn RL stage, fine-grained rewards gradually align preferences with the true user preferences even under biased feedback, enhancing robustness and generalization. A dual-view retriever combining semantic and collaborative signals ensures accurate candidate sets and prevents generating items outside the global item space.

\subsection{Ablation: Recommender and User Simulator}
\label{ex_RQ2}
To evaluate the impact of the SFT stage and the multi-turn RL stage on the recommender’s performance, we conducted ablation experiments on the INSPIRED dataset. The experimental settings are as follows:
\textbf{w/o RL}: The recommender trained with SFT only, without RL.
\textbf{w/o SFT}: The recommender trained with RL only, without SFT.
\textbf {w/o Pref}: The recommender’s preference inferences are not used during candidate retrieval.
\textbf{w/o Rec}: The recommender’s recommendation list is not used during candidate retrieval. As shown in Table~\ref{tab:RQ2}, removing SFT significantly degrades performance (e.g., Recall@1 drops from 0.099 to 0.056), highlighting its key role in warm-starting the model and enhancing reasoning with limited data. Removing RL also lowers performance (Recall@1 = 0.075), though less severely, indicating RL helps refine preference modeling and recommendation strategies via reward feedback. Ignoring preference (w/o Pref) or recommendation lists (w/o Rec) during retrieval reduces all metrics, showing the critical role of recommender–retrieval interaction.

\begin{figure}[t]
  \noindent
  \begin{subfigure}[b]{0.48\linewidth}
    \raggedright
    \includegraphics[width=\linewidth]{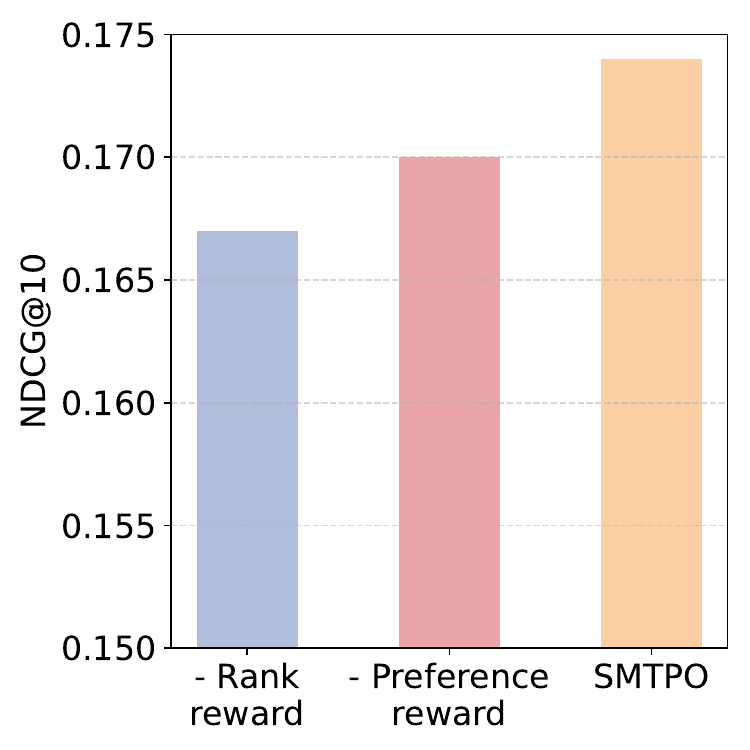}
    \caption{NDCG}
    \label{fig:RL_ndcg}
  \end{subfigure}
  \hfill
  \begin{subfigure}[b]{0.48\linewidth}
    \raggedleft
    \includegraphics[width=\linewidth]{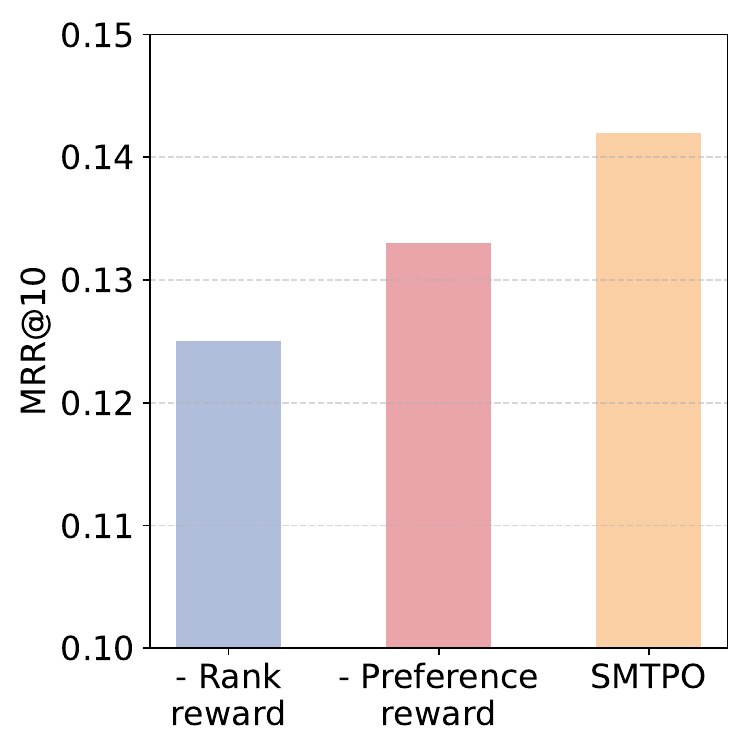}
    \caption{MRR}
    \label{fig:RL_mrr}
  \end{subfigure}
  \caption{The performance impact of different rewards on the INSPIRED dataset~\cite{hayati2020inspired}.}
  \label{fig:RL_rq2}
\end{figure}

To evaluate the effectiveness of reward design in the multi-turn RL stage, we conducted ablation studies on the reward design. Since the format reward ensures correct interaction with the simulator and retriever, we focused on the contributions of the recommendation and preference rewards. To retain basic recommendation ability, the hit reward was kept, while the preference and rank rewards were removed in experiments. As shown in Figure~\ref{fig:RL_rq2}, removing any of these rewards leads to a clear performance drop, with the rank reward having the greatest impact, highlighting its importance for the recommender’s reranking ability.

To evaluate the impact of multi-task SFT on the effectiveness of feedback generation, we conducted ablation studies on the simulator module of SMTPO, as shown in Table~\ref{tab:RQ2}. The settings are as follows:
\textbf{w/ Single-task}: Simulator is fine-tuned on a single task (feedback generation) instead of all tasks.
\textbf{w/o Feedback}: Natural language feedback from the simulator is not used during candidate generation. Results show that single-task SFT (w/ Single-task) underperforms full multi-task SFT (SMTPO) across all metrics, indicating that multi-task SFT enables the simulator to produce more detailed and informative feedback, better guiding the recommender. The simulator fine-tuned with single-task training (w/ Single-task) still outperforms most baselines, which indirectly demonstrates the effectiveness of our method in handling low-quality feedback. Furthermore, unused simulator feedback (w/o Feedback) during retrieval leads to further performance drops, especially in Recall@10 and NDCG@10, highlighting the critical role of natural language feedback in candidate selection.
\begin{table}[!t]
    \centering
    \caption{Ablation study of the simulator and the recommender on INSPIRED dataset. The best result is given in bold. Significant improvements are marked with * (t-test, p < 0.05).}
    \resizebox{\columnwidth}{!}{
    \begin{tabular}{lcccc}
        \toprule
        Method & Recall@1 & Recall@10 & NDCG@10 & MRR@10 \\
        \midrule
        w/o RL    & 0.075 & 0.203 & 0.129 & 0.107 \\
        w/o SFT   & 0.056 & 0.187 & 0.108 & 0.084 \\
        w/o Pref  & 0.082 & 0.217 & 0.144 & 0.122 \\
        w/o Rec   & 0.076 & 0.207 & 0.136 & 0.113 \\
        \midrule
        w/ Single-task & 0.095 & 0.220 & 0.152 & 0.130 \\
        w/o Feedback  & 0.072 & 0.197 & 0.126 & 0.105 \\
        \midrule
        SMTPO     & \bf{0.099*} & \bf{0.273*} & \bf{0.174*} & \bf{0.142*} \\
        \bottomrule
    \end{tabular}}
    \label{tab:RQ2}
\end{table}

\subsection{Multi-Turn Optimization}
\label{ex_RQ3}
\begin{figure}[t]
  \noindent
  \begin{subfigure}[b]{0.48\linewidth}
    \raggedright
    \includegraphics[width=\linewidth]{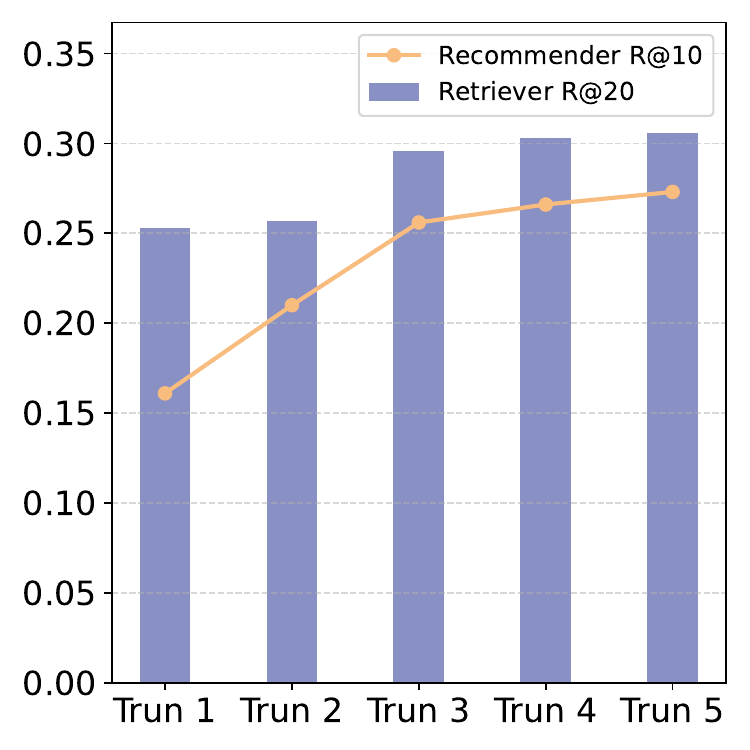}
    \caption{Recall}
    \label{fig:inspired_R}
  \end{subfigure}%
  \hfill
  \begin{subfigure}[b]{0.48\linewidth}
    \raggedleft
    \includegraphics[width=\linewidth]{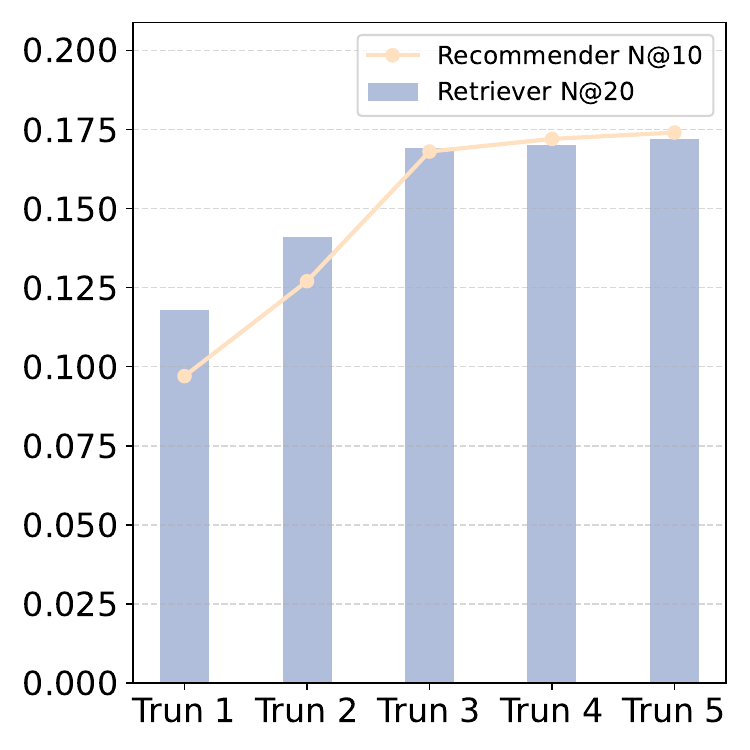}
    \caption{NDCG}
    \label{fig:inspired_N}
  \end{subfigure}
  \caption{Impact of multi-turn preference optimization on recommendation performance.}
  \label{fig:combined_rq2}
\end{figure}

To evaluate the effect of multi-turn preference optimization, we analyzed SMTPO’s recommendation performance over multiple conversation turns. Figures~\ref{fig:inspired_R} and~\ref{fig:inspired_N} show the recommender’s R@10/N@10 and the retriever’s R@20/N@20 across turns.
\begin{table}[t]
    \centering
    \caption{Blind-simulator and Noisy-feedback test on SMTPO. The best result is given in bold. Significant improvements are marked with * (t-test, p < 0.05).}
    \resizebox{\columnwidth}{!}{
    \begin{tabular}{lccccc}
    \toprule
    R@10 & Turn 1 & Turn 2 & Turn 3 & Turn 4 & Turn 5 \\
    \midrule
    Blind Simulator  & 0.161 & 0.194 & 0.171 & 0.164 & 0.167 \\
    Noise Test   & 0.161 & 0.182 & 0.185 & 0.178 & 0.181 \\
    SMTPO & \bf{0.161} & \bf{0.210*} & \bf{0.256*} & \bf{0.266*} & \bf{0.273*} \\
    \bottomrule
    \end{tabular}
    }
    \label{tab:RQ4}
\end{table}

Results indicate that recommender’s R@10 and N@10 steadily improve with more turns, demonstrating that multi-turn interactions help it better capture and refine user preferences. Notably, recommender’s N@10 eventually surpasses the retriever’s N@20, highlighting its effectiveness in reranking. In contrast, retriever’s R@20 and N@20, which mainly handle candidate generation, show limited improvement, indicating that the fine-ranking stage plays a more critical role in optimizing final recommendation performance. Overall, multi-turn preference optimization can gradually establish a reliable understanding of user preferences within a limited number of turns, effectively avoiding error accumulation caused by biased feedback, and thus significantly improving the accuracy and robustness of interactive conversational recommendation.

\begin{table*}[t]
\centering
\small
\caption{Case study of multi-turn preference optimization on the inspired dataset.}
\resizebox{\textwidth}{!}{
\begin{tabular}{|p{1.6cm}|p{15.9cm}|}
\hline
\multicolumn{2}{|c|}{\textbf{Base Info}} \\ \hline
\textbf{Target Item} &
\textcolor{red}{Joker} \quad
"category": [\textcolor{magenta}{drama film}, \textcolor{magenta}{psychological thriller film}, \textcolor{magenta}{crime film}];
\quad "director": \textcolor{magenta}{Todd Phillips}
\\ \hline

\textbf{History} &
\textbf{user:} I'm looking for something \textcolor{blue}{superhero}-related, like \emph{Guardians of the Galaxy}.  
\textbf{recommender:} Are you interested in comic-book characters?  
\textbf{user:} Yes, I love characters like \textcolor{blue}{Spider-Man}, \textcolor{blue}{Iron Man}, and \textcolor{blue}{Captain America}.  
\textbf{recommender:} Would you like to watch the trailer for \emph{The \textcolor{blue}{Avengers}}?  
\textbf{user:} Yes, that sounds great.  
\textbf{recommender:} That's the movie with \textcolor{blue}{Thanos}, right?  
\textbf{user:} Yes, \textcolor{blue}{Thanos} is in it.  
\textbf{recommender:} Superhero team-ups are always fun.  
\textbf{user:} Agreed, these \textcolor{blue}{superhero} mashups have been great.
\\ \hline

\multicolumn{2}{|c|}{\textbf{Turn 0 (Cold Start)}} \\ \hline

\textbf{Candidates(I)} &
Avengers(1), Avengers: Endgame(2), Marvel(3), ... \textcolor{red}{Joker}(15), ... Suicide Squad(20)
\\ \hline

\textbf{Reasoning(O)} &
Step1: Preference Inference. The user prefers \textcolor{blue}{action} \textcolor{blue}{superhero} movies, mentioning \textcolor{blue}{Guardians of the Galaxy}, \textcolor{blue}{Spider-Man}, \textcolor{blue}{Iron Man}, \textcolor{blue}{Captain America}, and the \textcolor{blue}{Avengers} with \textcolor{blue}{Thanos}, indicating a taste for \textcolor{magenta}{thrilling} films. Step2: …
\\ \hline

\textbf{RecList(O)} &
Avengers(1), Avengers: Endgame(2), Avengers: Infinity War(3), ... Aquaman(10)
\\ \hline

\multicolumn{2}{|c|}{\textbf{Turn 1 (Warm Up)}} \\ \hline
\textbf{Feedback(I)} &
I enjoy \textcolor{blue}{superhero} films featuring team-ups and epic battles against villains like \textcolor{blue}{Thanos}, especially in Marvel productions that blend \textcolor{blue}{action} with character depth, ensemble dynamics, and emotional stakes.
\\ \hline

\textbf{Candidates(I)} &
Star Wars(1), Avengers(2), Black Panther(3), ... \textcolor{red}{Joker}(8), ... Black Widow(20)
\\ \hline

\textbf{Reasoning(O)} &
Step1: Preference Inference. The user prefers intense, character-driven films with \textcolor{magenta}{psychological drama} and \textcolor{magenta}{crime} elements, emphasizing dark atmosphere, societal themes, and strong performances. Step2: …
\\ \hline

\textbf{RecList(O)} &
Blade Runner(1), Blade Runner 2049(2), \textcolor{red}{Joker}(3),... Avengers: Infinity War(10)
\\ \hline

\multicolumn{2}{|c|}{\textbf{Turn 2 (Warm Up)}} \\ \hline
\textbf{Feedback(I)} &
I enjoy \textcolor{blue}{superhero} team-up films with high stakes, mixing \textcolor{blue}{action}, sci-fi, and humor, featuring ensemble casts like \textcolor{blue}{Iron Man} and \textcolor{blue}{Captain America} against threats such as \textcolor{blue}{Thanos}, especially in Marvel productions.
\\ \hline

\textbf{Candidates(I)} &
Star Wars(1), Avengers(2), Black Panther(3), ... \textcolor{red}{Joker}(9), ... John Wick(20)
\\ \hline

\textbf{Reasoning(O)} &
Step1: Preference Inference. The user prefers \textcolor{magenta}{crime} \textcolor{magenta}{dramas} with \textcolor{magenta}{thriller} elements, focusing on complex characters, dark atmosphere, and strong performances, with an emphasis on \textcolor{magenta}{crime} themes. Step2: …
\\ \hline

\textbf{RecList(O)} &
\textcolor{red}{Joker}(1), Blade Runner 2049(3), Blade Runner(3), ... Star Wars(10)
\\ \hline
\end{tabular}
}
\label{tab:case_study}
\end{table*}

\subsection{Source of Performance Gains}
\label{ex_RQ4}
To analyze the sources of SMTPO's performance gains, we study the role of dynamic inputs in multi-turn interactions. The user simulator generates feedback based on both dialogue context and previous recommendations. To verify the necessity of dynamic inputs, we conduct a blind simulator experiment where the simulator only accesses dialogue history without recommender outputs. As shown in Table~\ref{tab:RQ4}, performance drops significantly, indicating that SMTPO’s gains come from dynamic interaction modeling rather than pretraining memorization.

We further examine whether the recommender truly learns from multi-turn feedback. To rule out gains from prompt engineering or model memorization, we conduct a noise injection experiment, randomly replacing simulator feedback with irrelevant chat. As shown in Table~\ref{tab:RQ4}, recommendation performance drops and shows no improvement over turns, indicating that SMTPO’s gains stem from continuously fitting real user preferences via feedback signals, rather than from prompt engineering or model memorization.

Overall, the multi-turn preference optimization mechanism gradually builds a reliable estimate of users’ implicit preferences within a few turns. During training, the recommender uses a multi-view reward to align dialogue context and simulator feedback with target items. At inference, it updates its understanding of user preferences from the learned feedback-driven posterior, enabling robust preference elicitation and improved recommendation performance.

\subsection{Case Study}
Table~\ref{tab:case_study} presents a representative example illustrating the effectiveness of SMTPO in multi-turn preference optimization. As the interaction progresses, the retriever consistently recalls the target item, and the majority of retrieved candidates share salient attributes with it. This observation highlights the robustness of our dual-view modeling strategy in fusing dynamic interaction signals across turns.

Meanwhile, the simulator-generated feedback at each turn is well aligned with the entity information in the dialogue context and provides targeted guidance for refining the current recommendation results. This demonstrates that multi-task instruction fine-tuning enables the simulator to produce specific and constructive user feedback rather than generic responses.

From the recommendation outcomes across turns, we observe that the recommender initially fails to select the target item from the candidate set. However, with continued multi-turn interactions, the target item is gradually ranked higher, eventually emerging as the top recommendation. This behavior indicates that the proposed multi-turn reinforcement learning framework allows the recommender to iteratively refine its understanding of user preferences and continuously improve recommendation quality. Moreover, the evolving preference summary increasingly aligns with the attribute information of the target item, further validating the effectiveness of the proposed multi-turn preference optimization mechanism.

\section{Conclusion}
This work presents SMTPO, a simulator-guided conversational recommendation framework based on Reasoning LLM that leverages a user simulator to enable multi-turn preference optimization. Within the framework, the simulator generates high-quality feedback via multi-task instruction tuning to guide the recommender in understanding complex user preferences. The recommender is trained in two stages, gradually aligning with true preferences over multiple interactions, while a dual-view dynamic retriever (semantic and collaborative) constrains the candidate set, enhancing precision and avoiding items outside the global item space. Experiments show that multi-task SFT significantly improves simulator feedback quality, and both SFT and multi-turn RL are crucial for the recommender’s preference modeling and recommendations. Multi-turn preference optimization allows the recommender to progressively build reliable user preference understanding. Overall, SMTPO effectively exploits LLM reasoning, achieving notable gains in accuracy, robustness, and generalization. Our work provides a potential direction for future exploration of applying large language models in multi-turn preference optimization.
\bibliographystyle{ACM-Reference-Format}
\bibliography{sample-base}

\appendix

\end{document}